\documentclass[twocolumn, preprintnumbers, 
endnote,nofootinbib,prl,9pt, superscriptaddress]{revtex4}

\usepackage{epsfig,subfigure}

\usepackage{epstopdf}

\usepackage[utf8]{inputenc}
\usepackage{graphicx}
\usepackage{amssymb}
\usepackage{amsxtra}
\usepackage{amsmath}
\usepackage{booktabs,multirow,tabularx}
\usepackage{slashed}
\usepackage{float}
\usepackage{placeins}
\usepackage{rotating}
\usepackage{lscape}
\usepackage{color}
\usepackage{hyperref}

\usepackage[Q=yes,pverb-linebreak=no]{examplep}

\DeclareGraphicsExtensions{.eps}
\graphicspath{{./}}

\newcommand{\vev}[1]{\langle {#1} \rangle}
\newcommand{\lsim}{\lesssim}
\newcommand{\gsim}{\gtrsim}
\newcommand{\eps}{\varepsilon}

\newcommand{\ord}[1]{\mathcal{O}{(#1)}}
\newcommand{\eq}[1]{Eq.~(\ref{#1})}

\newcommand{\ud}{U(1)_d}
\newcommand{\ad}{\alpha_d}
\newcommand{\mad}{m_{A_d}}

\def\beq{\begin{equation}}
\def\bea{\begin{eqnarray}}
\def\eeq{\end{equation}}
\def\eea{\end{eqnarray}}
\def\beqnl{\begin{align}}
\def\endal{\end{align}}

\definecolor{red1}{cmyk}{0,1,1,0.1}
\definecolor{blue1}{cmyk}{1,0,0,0}

\DeclareFontFamily{U}{cbgreek}{}
\DeclareFontShape{U}{cbgreek}{m}{n}{
        <-6>    grmn0500
        <6-7>   grmn0600
        <7-8>   grmn0700
        <8-9>   grmn0800
        <9-10>  grmn0900
        <10-12> grmn1000
        <12-17> grmn1200
        <17->   grmn1728
      }{}
\DeclareFontShape{U}{cbgreek}{bx}{n}{
        <-6>    grxn0500
        <6-7>   grxn0600
        <7-8>   grxn0700
        <8-9>   grxn0800
        <9-10>  grxn0900
        <10-12> grxn1000
        <12-17> grxn1200
        <17->   grxn1728
      }{}

\makeatletter
\newcommand{\normalorbold}{%
  \ifnum\pdf@strcmp{\math@version}{bold}=\z@ bx\else m\fi
}
\makeatother

\begin{document}

\date{\today}

\title{\boldmath GeV-Scale Messengers of Planck-Scale Dark Matter}

\author{Hooman Davoudiasl\footnote{email: hooman@bnl.gov}
}

\author{Gopolang Mohlabeng\footnote{email: gmohlabeng@bnl.gov}
}

\affiliation{Physics Department, Brookhaven National Laboratory,
Upton, New York 11973, USA}

\begin{abstract}

If dark matter (DM) originates from physics near the Planck scale it could be directly detected via its multiple scattering signals, yet this requires a large cross section for DM interactions with atoms.  Hence, detection of such DM could imply mediation by new low mass messengers.  We propose that a dark $U(1)_d$ remnant of the underlying spacetime geometry or a unified theory may survive down to small mass scales $\sim 1$~GeV, connecting low energy Standard Model (SM) and Planck scale phenomena.  Typical required cross sections for direct detection of Planck scale DM can be achieved through the $U(1)_d$ interactions of DM with SM quarks.  Low energy intense sources may uncover the GeV scale messengers of Planckian physics, allowing for testable predictions.  We assume that $U(1)_d$ is gauged baryon number, which implies several new electroweak charged particles are expected to arise near the weak scale to cancel gauge anomalies.  The model generically gives rise to kinetic mixing between the $U(1)_d$ gauge boson and the photon, which may be measurable.  In this scenario, direct detection of DM and measurements of a low energy messenger, including its kinetic mixing with the photon, can potentially shed light on the high energy character of the scenario.  Astrophysical considerations related to white dwarf stability against runaway nuclear fusion potentially disfavor DM heavier than $\sim 10^{17}$~GeV within our assumed messenger model.

\end{abstract}

\maketitle

\section{Introduction\label{sec:intro}}

While the presence of dark matter (DM) as a major component - about 25\% - of the cosmic energy budget has robust observational support, its basic properties remain largely unknown.  
In particular, possible values of DM mass cover a vast range of possibilities.  Given this situation, DM searches have often been guided by theoretical motivation.  A well-known example is weak scale DM, whose potential connection with electroweak physics has been a motivating factor to focus searches in this direction.  However, as experimental data further constrain physics at the weak scale this scenario becomes less compelling.  
This has motivated expanding DM searches away from the weak scale, mainly to lower masses.  
In particular, a substantial part of the parameter space remains to be probed in direct detection experiments for DM masses at or below the GeV scale, given the low kinetic energy of Galactic light DM and the weakness of the corresponding direct detection signals. 

One could in principle also move towards DM masses above the weak scale.  As DM mass gets larger  and its inferred number density drops, the constraints get less strong.  
Even though it is not clear where new physics emerges above the weak scale, the implied scale of quantum gravity, given by the 
Planck mass $M_{\rm P} \approx 1.2 \times 10^{19}$~GeV, offers an obvious target.  Somewhat below this scale one also gets motivation from string theory and Grand Unified Theories (GUTs), corresponding to masses $\gsim 10^{15}$~GeV.  
Hence, the range $\sim 10^{15-19}$~GeV provides an interesting target for the scale of DM physics, far above the weak scale.  We will refer to DM in this range of masses, generally speaking, as Planck Scale DM (PSDM).  

While PSDM may be well motivated from a theoretical point of view, it poses experimental challenges due to its extremely high mass, placing it entirely  outside the reach of accelerator or collider experiments.  
Assuming the usual local Galactic energy density for DM, $\rho \approx 0.3$~GeV/cm$^3$, it follows that the PSDM number density is very low within the Solar System.  
However, this also implies that the bounds on the interactions of PSDM with ordinary matter are not strong and this type of DM could potentially have a significant cross section for scattering off the target material. 
In fact, this possibility suggests that PSDM could scatter multiple times inside a detector and the ambient matter, unlike weak scale DM, giving rise to distinct signals \cite{Bramante:2018qbc}. Furthermore, multiple scattering of PSDM off material inside stars could have significant implications for stellar dynamics as pointed out for the case of white dwarfs by Ref.~\cite{Graham:2018efk}.  

Based on the results of Refs.~\cite{Bramante:2018qbc, Graham:2018efk}, depending on the detector type, a nucleon-DM cross section
$\sigma_{n\chi}\sim 10^{-36}-10^{-30}$~cm$^2$ provides an interesting range for searches by current and future envisioned DM detectors. 
However, the relatively large cross sections required for this purpose are typically not associated with such high mass scales. Therefore, one is led to consider light mediators that connect Planckian physics to low energy Standard Model (SM) states, such as quarks or leptons.

Vector bosons associated with gauge symmetries are well-motivated light particles, whose mass is protected by gauge invariance from receiving large quantum corrections.  To keep the vector boson 
light compared to Planckian mass scales, one then needs to consider gauge symmetries that are not broken ``easily.''  Interestingly, it turns out that the simplest gauge group, namely a $U(1)$ enjoys such resilience and is typically resistant to spontaneous symmetry breaking.  
Also, $U(1)$ factors are quite ubiquitous in string theory 
compactifications and as by-products of symmetry breaking in GUTs; 
see for example Ref.~\cite{Halverson:2018xge}.  Hence, a $U(1)$ gauge symmetry can naturally be assumed to provide a connection between Planckian physics and low energy dynamics.     

In this work, we consider vector bosons $A_d$ of mass $\mad$ associated with a ``dark'' gauged $\ud$ as the aforementioned light messenger that connects the PSDM and ordinary matter sectors. 
The $\ud$ coupling constant will be denoted by $g_d$.  For concreteness, it will be assumed  that PSDM is a Dirac fermion $\chi$ of mass $m_\chi$ and that $A_d$ couples to SM quarks. This may be interpreted as a coupling to baryon number \cite{Rajpoot:1989jb,Nelson:1989fx,He:1989mi,Foot:1989ts,Bailey:1994qv,Carone:1994aa,Aranda:1998fr,FileviezPerez:2010gw,FileviezPerez:2011pt,Graesser:2011vj,Pospelov:2011ha,Pospelov:2012gm,Dobrescu:2014fca,Tulin:2014tya,Batell:2014yra,Perez:2014qfa,Coloma:2015pih,Perez:2015rza}, which we will implicitly adopt for the rest of our work. 
Other charge assignments could potentially lead to viable models. 
However, we note that it may not be straightforward to construct a model, based on other gauged symmetries, that can result in sufficiently large scattering cross sections for multiple scattering of PSDM from matter.  One can also entertain the possibility that the interactions between PSDM and the SM are mediated by a light scalar, however we will not focus on this possibility.  In the remainder of this work, we will introduce our model and its properties, discuss the current constraints on the model parameter space as well as
provide possible implications for future experiments.

\section{Model Description  \label{sec:mod}}

The Lagrangian for the interactions of $A_d$ can be written as 
\beq
{\cal L}_d = g_d (Q^q_d\, \bar q \gamma_\mu q + Q^\chi_d\, \bar \chi \gamma_\mu \chi) A_d^\mu         
\,,
\label{Ld}
\eeq
where $Q^q_d = 1/3$ is the $\ud$ charge (baryon number) of SM quarks denoted by $q$; 
we will assume $Q^\chi_d = 1/2$ which is consistent with $\chi$ stability as a DM candidate, if it is the lightest state with this $\ud$ charge.  

As presented here, the $U(1)_d$ is anomalous and hence does not represent a consistent gauge theory \cite{Bardeen:1969md}.  
In order to cancel the baryon current-electroweak $SU(2)\times U(1)_Y$ anomalies, we will need to posit that new fermions $F_i$, $i=1,2,3,\ldots, n$,  with electroweak quantum numbers and chiral charges under $\ud$ appear in the ultraviolet regime \cite{FileviezPerez:2010gw, Perez:2015rza}. 
Given the assumed charges, these fermions must have masses $m_F\gsim 100$~GeV to have evaded detection in high energy experiments, so far.  The chiral nature of these fermions suggests that they get their masses from a Higgs field $\Phi$ charged under $\ud$.  

The ultraviolet model Yukawa couplings that generate masses for $F_i$ are of the form 
\beq
y\, \Phi \bar F_L F_R.  
\label{Yukawa}
\eeq

To get $m_F\gsim 100$~GeV, required by experimental constraints, and for $y\lsim 1$, we then need a vacuum expectation value $\vev{\Phi}\gsim 100$~GeV.  However, as we will discuss below, for typical parameters in our scenario $\mad \sim 1$~GeV and $g_d \lsim e$, where $e$ is the electromagnetic coupling. 
Since the mass of $A_d$ is given by $Q^\Phi_d \, g_d \vev{\Phi}$, we would need $\vev{\Phi}\sim 1$~GeV if the $\ud$ charge $Q^\Phi_d$ of 
$\Phi$ is $\ord{1}$.  Such a small value for $\vev{\Phi}$ would lead to GeV scale $F_i$, in severe conflict with experimental data \cite{Dobrescu:2014fca, Halverson:2014nwa, Heister:2002mn}.   
We thus conclude that we need $Q^\Phi_d \ll 1$ in order to have $\vev{\Phi}\gsim 100$~GeV.   
Consequently, \eq{Yukawa} implies that the chiral charges of $F_i$ are $Q^F_d \ll 1$ also.  Therefore, to cancel the $\ud$ anomalies caused by $\ord{1}$ SM charges we would need many $F_i$, {\it i.e.} $n\gg 1$.  In order to avoid requiring 
exceedingly large values of $n$, $\vev{\Phi}$ cannot be too large, while maintaining $\mad\sim 1$~GeV \cite{Dobrescu:2014fca}.  

Let us also add that if the charge of PSDM is sufficiently larger than assumed above, the same values of DM-SM scattering cross sections can be achieved for 
smaller values of $\ad$.  This could allow for $\mad \sim 1$~GeV as assumed below, without resulting in values of $m_F$ that are too small.  This could potentially be an alternate way to avoid bounds on anomaly canceling fermions, while keeping $\mad$ at the GeV scale. 

Thus, we would generically expect the appearance of several electroweak charged fermions $F_i$ not far above the weak scale in our scenario. For example, the typical models considered in Ref.~\cite{Dobrescu:2014fca} suggest that for $\ad \lsim 10^{-2}$, the required number of new heavy fermions of mass $\gsim 100$~GeV is $n\lsim 20$.  According to the results of Ref.~\cite{Preskill:1990fr}, the bound on the effective field theory 
cutoff scale of the anomalous $U(1)_d$ is $\Lambda_{\rm cutoff} \lsim (4\, \pi)^3 \mad/(g_{\rm EW}^2 \,g_d)$ where $g_{\rm EW}$ represents an electroweak gauge coupling (see also  Refs.~\cite{Dror:2017ehi,Dror:2017nsg,Tulin:2014tya}).  For $\mad \sim 1$~GeV and $\ad \lsim 10^{-2}$, as considered in our work, one then sees that the maximum cutoff scale could be $\Lambda_{\rm cutoff}\sim 20$~TeV or larger, consistent with the new fermions appearing at or above $\sim 100$~GeV.

Interestingly, the nature of the above anomaly-canceling fermions leaves an imprint on low energy phenomena, which could lead to significant effects.  In particular, if these fermions get their mass from electroweak symmetry (EWS) preserving sources, corresponding to a $\Phi$ that is a singlet under the SM gauge symmetries, low energy processes get a non-decoupling contribution from the longitudinal mode of $A_d$ \cite{Preskill:1990fr, DHoker:1984izu,DHoker:1984mif,Tulin:2014tya,Dror:2017ehi,Dror:2017nsg}.  We will assume that $m_F\neq 0$ is 
EWS preserving, to avoid further complications that arise if new sources of EWS breaking are present beyond the minimal SM Higgs.  However, as Refs.~\cite{Dror:2017ehi,Dror:2017nsg} have shown, new constraints beyond those considered in previous studies apply, {\it e.g.} from $Z$ boson and $B$ meson decays.

We note that the above interactions generically give rise to kinetic mixing of $A_d$ and the photon $\gamma$, through 1-loop effects \cite{Holdom:1985ag} \footnote{Our model may also accommodate an interaction between the SM Higgs and the scalar $\Phi$ introduced above. However for a weak scale $\vev{\Phi}$, we would require significant fine-tuning to obtain a GeV-scale $\Phi$. Without such a fine-tuning, one can show that the mixing with the SM Higgs will be of order $\lesssim 10^{-4}$. Hence, we do not pursue this possibility further. For further details see for example Ref.~\cite{Gopalakrishna:2008dv}. }. 
This allows $A_d$ to couple to the electromagnetic current $J_\mu^{\rm em}$ via $e \, \eps\, J_\mu^{\rm em} A_d^\mu$.    
The resulting loop-induced mixing parameter $\eps \ll g_d$ would not affect our conclusions regarding direct detection of PSDM significantly and hence would be ignored in that discussion.  

Let us examine the possible size of kinetic 
mixing parameter $\eps$.   All quarks couple to both 
hypercharge $U(1)_Y$ and $\ud$, thus they can mediate loop-level kinetic mixing.  To cancel anomalies, we also generally need fermions that couple to both $U(1)$ gauge fields.  In principle, 
$\eps$ is a renormalized parameter and as such its value can only be determined by measurement.  However, depending on the contributions from ultraviolet states of mass $M$, carrying 
both $U(1)_Y$ and $\ud$ charges, a typical estimate could be 
\beq
\eps \sim \frac{e\, g_d }{16 \pi^2} \ln \left (\frac{M}{\bar m} \right)
\label{eps-est}
\eeq 
where $\bar m\sim$~GeV represents a mean mass scale for the SM quarks.  For $M\sim m_\chi$ and $g_d \lsim 0.1$, we may then expect that values of $\eps \lsim 10^{-2}$ could be potentially achieved in our scenario.  However, experimental constraints may disfavor the upper end of this range depending on the value of $\mad$, as will be discussed below.  If states at scales of $\ord{m_\chi}$ do not contribute to $\eps$, we may then assume $M\sim m_F$ and hence 
$\eps \lsim 10^{-3}$.  We will not use a specific value of $\eps$ in what follows and the discussion above is only meant to provide a rough range of possibilities.

\section{Current Bounds and Results \label{sec:cons}}

\begin{figure*}[t]
\centering
\includegraphics[scale=0.55]{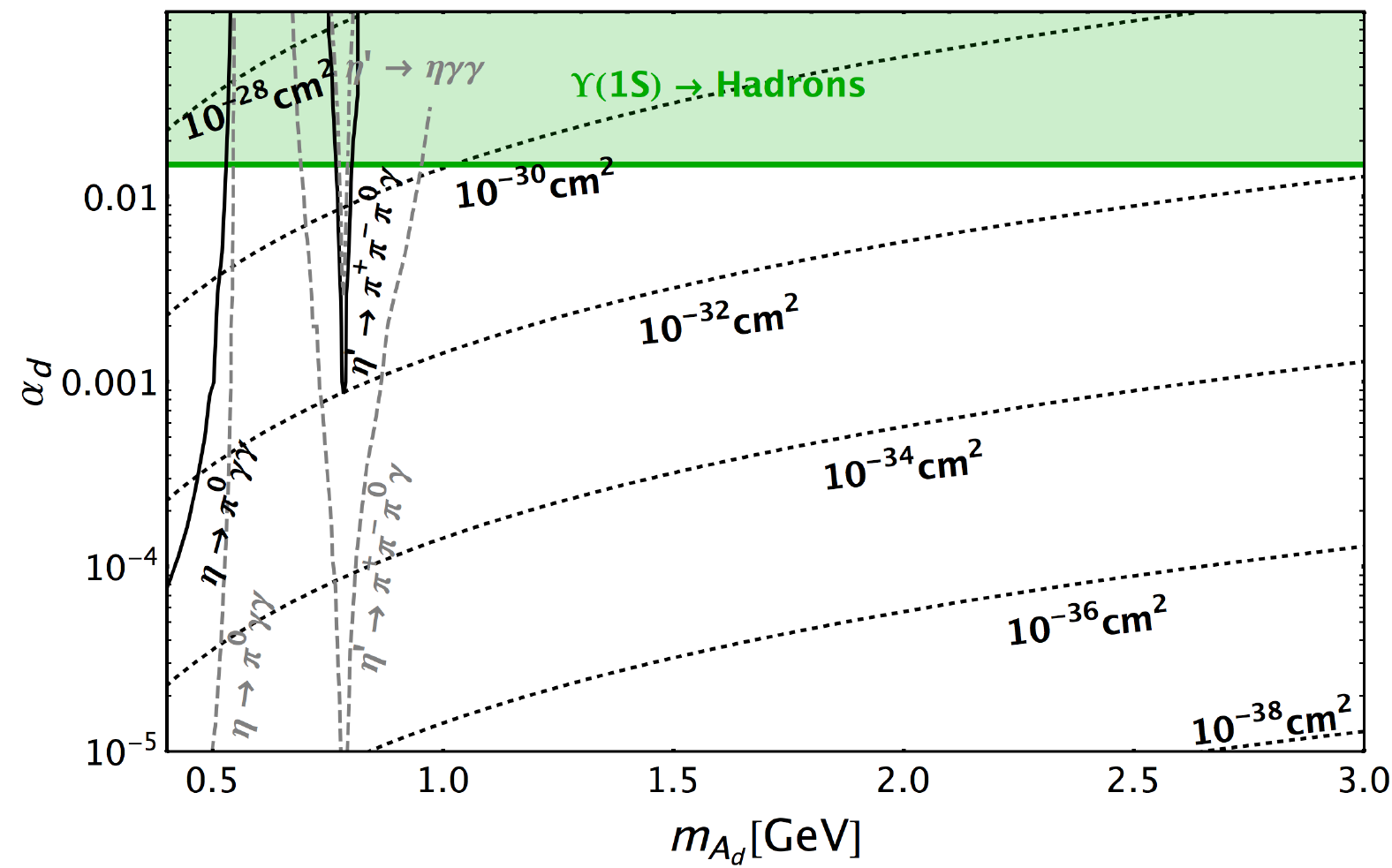}
\includegraphics[scale=0.55]{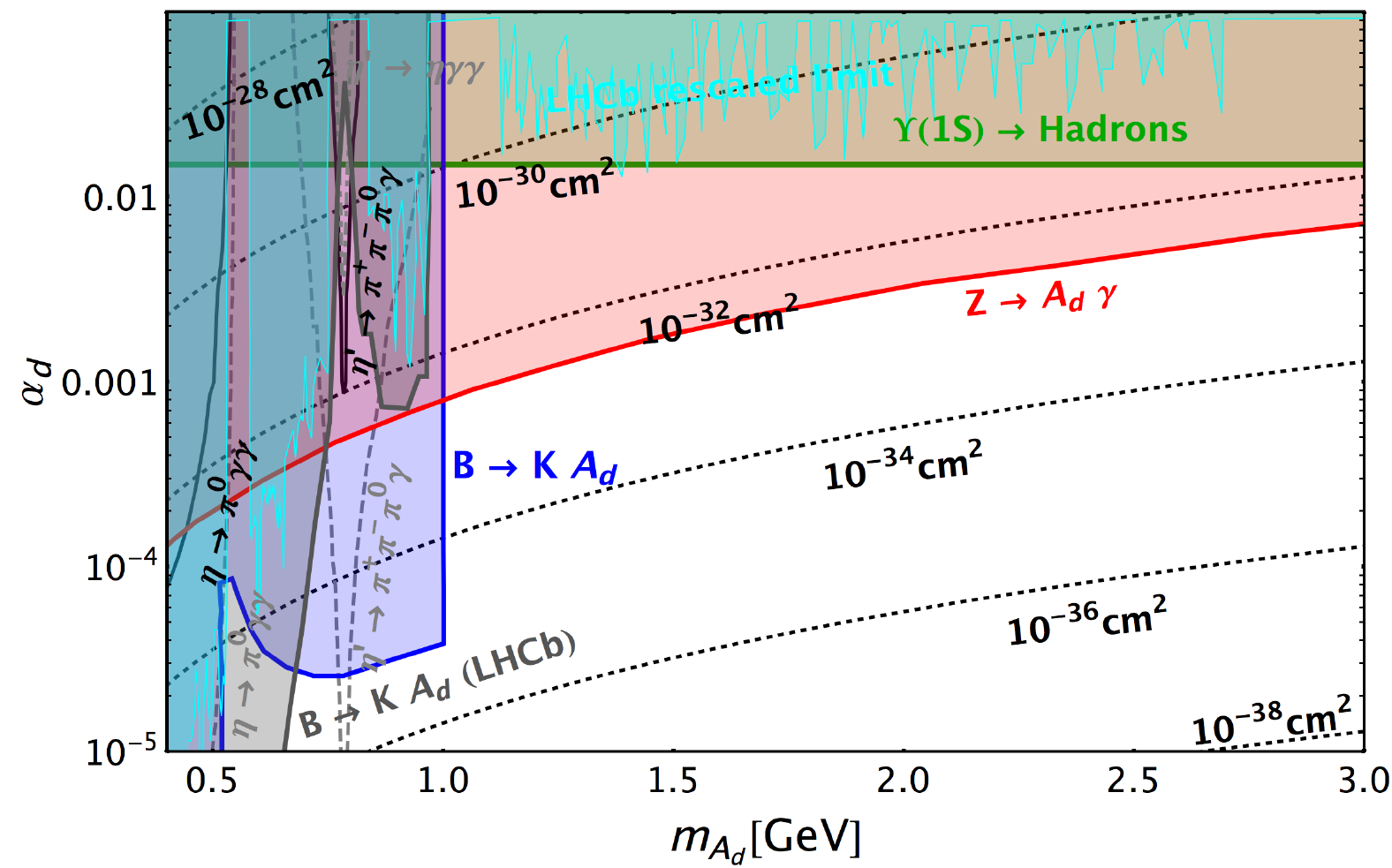}
\caption{Current limits on $\ad$, plotted versus $\mad$.  The dotted contours correspond to values of constant
$\sigma_{n \chi}$.   
The top green shaded region is ruled out by $\Upsilon (1S)$ decay, requiring $\alpha_{d} ~\textless $ 0.014 \cite{Carone:1994aa, Aranda:1998fr}.  
The solid black and gray dashed contours represent bounds and future projections obtained from anomalous $\eta$ and $\eta^{\prime}$ decays, as described in Ref.~\cite{Tulin:2014tya}.  The left panel shows bounds assuming the existence of an ultraviolet complete model without 
(energy/mass)$^2$ enhanced contributions from a longitudinal $A_d$ mode.  
The presented bounds are not sensitive to the levels of kinetic mixing considered in this work.
On the right panel we include bounds from enhanced longitudinal $A_d$ emission. The blue shaded region is the bound from $B \rightarrow K ~A_d$ and the red shaded region is the region constrained by $Z \rightarrow \gamma ~A_d$ as in Refs.~\cite{Dror:2017ehi,Dror:2017nsg}. The darker grey region illustrates bounds from $B \rightarrow K ~A_d$ searches at LHCb and are obtained from Ref.~\cite{Ilten:2018crw}. For the bounds in the red and grey regions $A_d$ is assumed to decay visibly to leptons, mediated by kinetic mixing. The bound in cyan is the LHCb limit on the search for visibly decaying dark photons rescaled to include a baryon coupling as in Ref.~\cite{Ilten:2018crw}. }
\label{alph_md}
\end{figure*}

In this section we will consider various experimental constraints that apply to our model.  We will adopt a parameter space allowed by the phenomenological analysis presented in Ref.~\cite{Tulin:2014tya} and we will largely focus on vector boson masses in the range $500 {\rm ~MeV} \lsim \mad\lsim 3$~GeV. We do not consider vector masses far below ~ GeV in order to avoid significant constraints from low energy hadron physics as is illustrated in Fig.~\ref{alph_md}. We will assume values of $\ad \lsim 10^{-2}$, where $\ad \equiv g_d^2/(4 \pi)$. For our Galactic neighborhood, 
the virial velocity $v\sim 10^{-3}$.  Given that $m_\chi$ is assumed 
to be many orders of magnitude heavier than nucleons of mass $m_n \sim$~GeV, the momentum 
transfer $q$ in nucleon-PSDM scattering is roughly given by $q\sim m_n v \lsim 1$~MeV.  Hence, it is safe to assume that $\mad \gg q$ in our treatment.  The spin-independent nucleon-$\chi$ scattering cross section can then be approximated by (see, for example, Refs.~\cite{Pospelov:2007mp,Essig:2011nj}) 
\beq
\sigma_{n \chi} \approx \frac{16 \pi \mu_{n \chi}^2\,(Q^n_d Q^\chi_d)^2 \ad^2}
{\mad^4}\,,
\label{sig-n-chi}
\eeq 
where the reduced mass $\mu_{n \chi} \approx m_n$ to a very good approximation in the case 
of PSDM and the $\ud$ (baryon) charge of a nucleon is $Q^n_d = 1$.  As discussed before, for $m_\chi$ values $\sim 10^{15-19}$~GeV of interest in our work, 
we could potentially expect $\eps \lsim 10^{-2}$.  Since we will only consider $\ad > 10^{-5}$ in what follows, it is a good approximation to ignore the effect of $e\,\eps$ compared to $g_d$, for the purposes of the above direct detection cross section estimate.  

In Fig.~\ref{alph_md}, we plot the values of $\ad$ versus $\mad$, where we illustrate the values corresponding to $\sigma_{n \chi} = 10^{-x}$~cm$^2$, for $x=28, 30,\ldots, 38$ (black dotted lines).  The green shaded region takes into account the bounds from $\psi$ and $\Upsilon$ quarkonium state decays to hadrons, the most stringent in our case being the $\Upsilon (1S)$ decay bound requiring $\alpha_{d} ~\textless $ 0.014 \cite{Carone:1994aa, Aranda:1998fr}.
The solid black and gray dashed contours represent bounds and future projections obtained from the search for anomalous decays of the pseudo-scalar mesons $\eta$ and $\eta^{\prime}$ by various experimental searches.
We refer the reader to Ref.~\cite{Tulin:2014tya} whose detailed analysis of these bounds and projections we utilize here. \footnote{We do not specify the constraints implied by the lower bound on the masses of the fermions that cancel the baryon number anomaly.  These bounds can have significant dependence on the details of the weak scale model building and hence do not provide concrete limits for the effective theory treatment in this work.  For a more detailed discussion, please see Refs.~\cite{Dobrescu:2014fca,Dror:2017ehi,Dror:2017nsg}.}

For phenomenological purposes, in this work we will not specify the exact ultraviolet completion that may satisfy the model description above. We have instead, as stated above, assumed the existence of heavy anomaly canceling fermions whose masses are obtained from EWS preserving sources. This leads to enhanced longitudinal $A_d$ mode contributions to low energy processes.  
However, one could in principle imagine an ultraviolet completion in which the longitudinal mode may not be enhanced \cite{Dror:2017ehi,Dror:2017nsg}.  In such a scenario, the low energy constraints, {\it e.g.} from $Z$ boson or $B$ meson decays, could be significantly weaker. We present this scenario on the left panel of Fig.~\ref{alph_md}, where the only bounds are from possible tree level interactions of the transverse $A_d$ with pseudo-scalar mesons. These bounds can receive negligible variations from the small kinetic mixing parameter $\eps$, for the range of parameters considered here.  

On the right panel of Fig.~\ref{alph_md}, we include the bounds from $Z \rightarrow \gamma ~A_d$ (red shaded region) and $B \rightarrow K ~A_d$ (blue shaded region), corresponding to longitudinal $A_d$ enhancement, assuming $\eps = e g_{d}/(4 \pi)^{2}$; please see Refs.~\cite{Dror:2017ehi,Dror:2017nsg} for more details regarding these bounds.  
We also include a bound from $B \rightarrow K ~A_d ~(A_{d} \to \mu^{+} \mu^{-})$ at the LHCb experiment \cite{Aaij:2015tna, Aaij:2016qsm}. These bounds are represented by the grey shaded region. In the cyan shaded region we take into account the constraints on visibly decaying dark photon searches, using leptonic final states, at LHCb (from Drell-Yan production), rescaled to incorporate a baryon coupling.  
Both LHCb bounds are obtained from Ref.~\cite{Ilten:2018crw} assuming $\eps = e g_{d}/(4 \pi)^{2}$. 
We note here that the cyan bounds are independent of any longitudinal mode enhancements, however since they assume a relation between $\eps$ and $g_d$ we include them only on the right panel and assume no relation between the two couplings on the left panel.
We see that, for $\mad \sim 1$ GeV, we can obtain values of $\sigma_{n \chi}$ needed for multiple or single scatterings in underground detectors \cite{Bramante:2018qbc}.  Here, we also add that those cross sections could be obtained at smaller values of $\ad$, in case PSDM has a $\ud$ charge $Q_d^\chi \gsim 1$, as may be the case for a composite state.

\begin{figure}[t]
\centering
\includegraphics[scale=0.54]{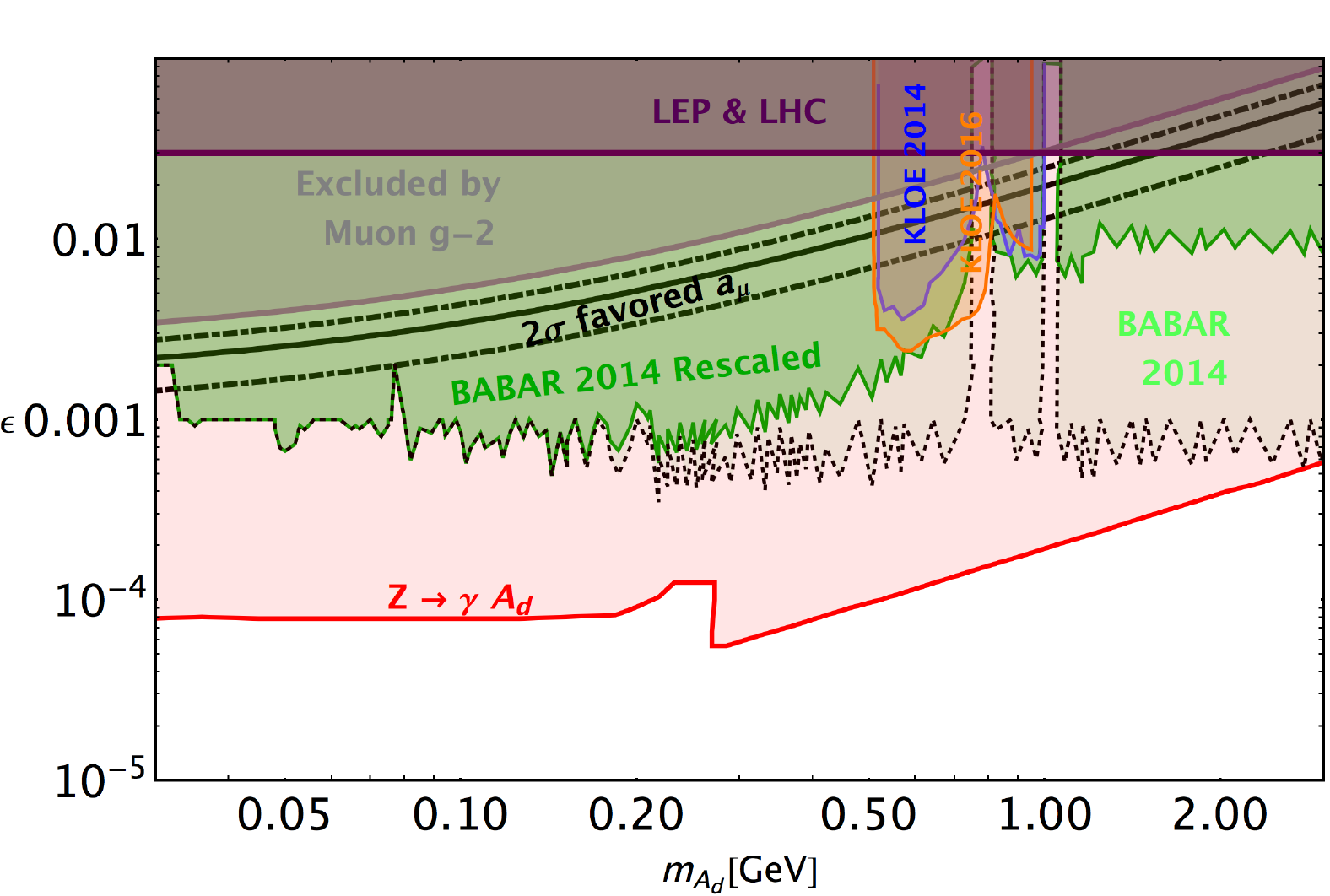}
\caption{Limits on the kinetic mixing parameter $\eps$ versus $\mad$. 
As an example we show the red shaded region which represents a bound from 
$Z \rightarrow \gamma ~A_d ~(A_d \rightarrow l^{+} l^{-})$ in the case when anomaly cancellation results in a longitudinal mode enhancement, assuming $\eps = e g_{d}/(4\pi)^{2}$ \cite{Dror:2017ehi,Dror:2017nsg}. 
For other limits we assume no longitudinal mode enhancement, treat $\eps$ as a free parameter, and set $\ad = 10^{-3}$. The light green shaded region (bounded by the black dotted line) is the limit from the BABAR collaboration on a ``dark photon'' decaying to $e^+e^-, \mu^+\mu^-$ \cite{Lees:2014xha}. The purple upper bound at $\eps = 3 \times 10^{-2}$ is the model independent limit from electroweak precision observables \cite{Hook:2010tw, Curtin:2014cca}. The grey shaded region is the 5$\sigma$ exclusion limit from the muon $g-2$ experiment and the black dotdashed band represents the 2$\sigma$ favored parameter space for the muon $g-2$ anomaly with the central value represented by the black solid line \cite{Bennett:2006fi, Pospelov:2008zw, Patrignani:2016xqp}. The darker green shaded region is the BABAR dark photon limit rescaled to include a baryon coupling. The blue shaded area represents a rescaled limit for dark photon decaying into a lepton pair by the KLOE 2014 collaboration and the orange shaded area is the rescaled limit for dark photon decaying into a pair of pions by the KLOE 2016 collaboration.}
\label{eps_md}
\end{figure}

In Fig.~\ref{eps_md}, we show the existing bounds on the visible ``dark photon'' kinetic mixing parameter, also denoted by $\eps$, versus $\mad$ in an ultraviolet model with anomaly free currents.  
This figure incorporates two alternate assumptions: (1) that the anomaly cancellation results in a longitudinal enhancement in which our model gets constraints from $Z$ boson decays and (2) that there is an ultraviolet completion in which we may cancel the gauge anomalies without having non conserved currents and longitudinal polarization enhancement at low energies.
The red shaded area in Fig.~\ref{eps_md} is the $Z \rightarrow \gamma A_{d}$ bound that arises from the presence of a longitudinal enhancement of the $A_{d}$, corresponding to case (1) above. Refs.~\cite{Dror:2017ehi,Dror:2017nsg} derived this bound on $g_d$ for $A_{d} \rightarrow l^{+} l^{-}$ in the parameter space represented here, assuming $\eps = e g_{d}/(4\pi)^2$, as a typical one-loop value.  We have used the assumed form of $\eps$ in Refs.~\cite{Dror:2017ehi,Dror:2017nsg} 
and mapped their bound on $g_d$ to $\eps$.  
It is difficult to map the $B$ meson bounds on this parameter space since they were derived assuming $A_{d} \to$ hadrons for $m_{\pi} \lsim \mad \lsim 1$ GeV. In principle one may also map out the cyan and grey regions in the right panel of Fig.~\ref{alph_md} assuming $\eps = e g_{d}/(4\pi)^2$, however for clarity of presentation we illustrate only the $Z$ boson bounds from possible longitudinal mode enhancements. 

In case (2), without longitudinal enhancement, we assume $\eps$ and $\ad$ are both free parameters of the model. Hence, we choose a benchmark value of $\ad = 10^{-3}$. 
In this scenario, dominance of hadronic decays of $A_d$ could make the constraints on the dark photon, derived from its assumed leptonic decays, much weaker.  
To see this, note that in the limit where the hadronic width of $A_d$ is approximated by the width into quarks, the $A_d$ branching fraction into leptons $(e, \mu)$ is suppressed by $\sim (e\,\eps/Q^q_d \,g_d)^2 \lsim 10^{-2}$, and hence dark photon constraints cannot be assumed for $A_d$.  
The semi-transparent green shaded region in Fig.~\ref{eps_md} is the parameter space ruled out by the BABAR experiment  in the search for a dark photon decaying into leptons $l=e,\mu$ \cite{Lees:2014xha}. The darker green shaded region represents a rescaled limit assuming a dominant baryon coupling relevant to the model discussed here.  

We obtain the above rescaled limit by using the simple form for the number of signal events as 
\beq
N_{A_d} = \sigma_{A_{d} \gamma}\, {\rm Br}(A_{d} \rightarrow l^{+} l^{-}) \,\mathcal{L},
\label{nad}
\eeq
where $N_{A_{d}}$ is the number of $A_{d}$ signal events, $\sigma_{A_{d} \gamma}$ is the production cross section for $A_{d}$ in association with a photon in $e^{+} e^{-}$ collisions,  ${\rm Br}(A_{d} \rightarrow l^{+} l^{-})$ is the branching ratio of $A_{d}$ into leptons and $\mathcal{L}$ is the integrated luminosity of the experiment. 
In our analysis, we require the same number of events as in the dark photon analysis, {\it i.e.} $N_{A_{d}} ~\approx ~N_{\rm BABAR}$. Using this assumption, we calculate the limit on $\eps$ in the $A_{d} \rightarrow l^{+} l^{-}$ mode by rescaling the values in the BABAR analysis with branching ratios to leptons after including the dominant decays into hadrons, for the case of the model adopted here.
We perform a similar procedure for the KLOE 2014 analysis, based on their search for a dark photon decaying predominantly into leptons \cite{Babusci:2014sta}, represented by the blue shaded region. 
In addition, the KLOE collaboration recently performed a search for a dark photon decaying predominantly into a pair of pions \cite{Anastasi:2016ktq}. 
For our study, we recast this limit in a similar manner as for the leptonic decay channel above. We assume a similar form as Eq.~(\ref{nad}) for the signal number of events, however we replace ${\rm Br}(A_{d} \rightarrow l^{+} l^{-})$ with the branching ratio of $A_d$ to pions, {\it i.e.} ${\rm Br}(A_{d} \rightarrow \pi^{+} \pi^{-})$. 
The recast KLOE region is illustrated by the orange shaded region labeled KLOE 2016. 
The Belle II experiment is expected to have an integrated luminosity of 50 $\rm ab^{-1}$ \cite{Abe:2010gxa, Kou:2018nap}. With this luminosity, Belle II could be able to probe an order of magnitude or more below the BABAR rescaled limit in $\eps$, depending on backgrounds and systematic uncertainties.

In the lower mass region, below the pion mass threshold, the branching ratio of the dark photon to leptons is more dominant both when assuming only kinetic mixing as well as in the presence of a baryon coupling, for the values we have considered here. 
Hence the two limits are the same, as is illustrated by the two green shaded regions.
Following the argument in Ref. ~\cite{Tulin:2014tya} we assign $A_d$ the same quantum numbers as the $\omega$ meson \cite{Patrignani:2016xqp}, which we use as a guide on how we expect $A_d$ to decay. 
For $m_{\pi} \lesssim \mad \lesssim 620$ MeV, $A_{d} \rightarrow \pi^{0} \gamma$ is the dominant decay mode, which we determine by assuming $A_d - \omega$ mixing. Beyond 620 MeV the dominant decay mode becomes $A_{d} \rightarrow \pi^{0} \pi^{+} \pi^{-}$. 
We simply assume here that the 3 pion decay mode is dominant until $4 \pi f_{\pi}$, which is taken to be the scale at which chiral perturbation theory breaks down, where $f_{\pi} = 93$ MeV is the pion decay constant \cite{Patrignani:2016xqp}. Beyond this scale we assume that $A_{d}$ decays directly into partons. 
In principle, near 1 GeV $A_{d}$ would decay like the $\phi$-meson, since it has similar quantum numbers. In this region, the decay rates of $A_d$ to 3 pions and to 2 kaons are roughly similar and this would not alter our conclusions significantly.

Since beyond the $3 m_{\pi}$ limit, both branching ratios of $A_{d} \rightarrow l^{+} l^{-}$ and $A_{d} \rightarrow \pi^{+} \pi^{-}$ are highly suppressed  
compared to $A_{d} \rightarrow \pi^{0} \pi^{+} \pi^{-}$ and $A_{d} \rightarrow \pi^{0} \gamma$, the corresponding BABAR and KLOE bounds on $\eps$ are loosened by a factor of $\ord{10}$ or more. 
Hence, for $\mad \gsim 800$~MeV we roughly end up with a bound of $\eps \lsim 10^{-2}$ for $\ad = 10^{-3}$.  
Also in Fig.~\ref{eps_md} we show the upper bound extracted from electroweak precision observables at LEP and LHC \cite{Hook:2010tw, Curtin:2014cca}. The gray shaded area is the region excluded by the muon $g-2$ experiment at 5$\sigma$ and the black dot-dashed band is the 
2$\sigma$ allowed explanation of the $g_{\mu}-2$ anomaly with the black solid line as the central value \cite{Bennett:2006fi, Pospelov:2008zw, Patrignani:2016xqp}.  
Absent longitudinally enhanced $A_d$ emission constraints, future measurements, for example by  
Belle II, should be able to probe the $\eps \lsim 10^{-2}$ region in Fig.~\ref{eps_md} in our setup.  In case a signal is detected in this regime, one could potentially conclude that the ultraviolet 
theory does not give rise to the longitudinal enhancements indicated by the red shaded region in the figure.  Hence, probing the kinetic mixing parameter, within the baryon current $\ud$ model, could in principle shed light on the underlying dynamics of anomaly cancellation at much higher energies.

\section{Astrophysical Considerations}

In Ref.~\cite{Graham:2018efk}, limits on the annihilation cross section of PSDM into SM states have been derived, based on the stability of white dwarfs against runaway nuclear fusion which would lead to a type Ia supernova.  The authors of Ref.~\cite{Graham:2018efk} find that the typical minimum mass for the PSDM trapped inside a white dwarf that would lead to a constraint is $\sim 10^{17}$~GeV.  

Given that we have adopted a specific model here, we should ensure that choices of parameters that could lead to potential direct detection of PSDM are consistent with astrophysical observations.  The pair annihilation cross section $\sigma_{\chi\chi} \, v_\chi$ of PSDM, $\bar\chi \chi \to A_d \, A_d$, in our scenario can be approximated by
\beq
\sigma_{\chi\chi} \, v_\chi \sim \frac{4 \pi \, \ad^2}{m_\chi^2} \sim 10^{-54}
\frac{\text{cm}^3}{\text{s}} \left(\frac{\ad}{10^{-2}}
\frac{10^{17}\,\text{GeV}}{m_\chi}\right)^2
\label{sigxx}
\eeq 
where $v_\chi$ is the typical velocity of $\chi$.  

The analysis in Ref.~\cite{Graham:2018efk} suggests that for $ 10^{17}\,\text{GeV} \lsim m_\chi \lsim 10^{19}\,\text{GeV}$, agreement with astrophysical observations require $\sigma_{\chi\chi} \, v_\chi \lsim 10^{-65}$~cm$^3$ s$^{-1}$, assuming a stable radius $r_c = 10^{-10}$~cm to which the DM has collapsed inside the star, $\sigma_{\chi n} = 10^{-32}$~cm$^2$, and $\rho_\chi = 0.4$~GeV cm$^{-3}$ as the local DM energy density.  Hence, within the model adopted in our work, the mass range $\sim 10^{17-19}$~GeV appears 
disfavored due to astrophysical constraints.\footnote{Those constraints also disfavor larger $m_\chi$ which may be interesting to consider, for example 
if DM is a composite state.}  
Possible deviations from the assumed parameters in that analysis, for example a larger value of $r_c$, could change the bounds and allow $m_\chi \gsim 10^{17}$~GeV to be viable within the model we have considered.

\section{Summary and Discussion\label{sec:disc}}

In this work we considered the possibility that the mass of dark matter 
may be close to the Planck scale $M_{\rm P}$ and lie in the range $\sim 10^{15-19}$~GeV.  In such a scenario, the direct detection bounds on the the cross section of dark matter interactions with 
ordinary matter, {\it e.g.} nucleons, are much weaker than for the typical weak scale models, due to the tiny implied number density of dark matter particles near the solar system.  Interestingly, the allowed large cross section could then lead to distinct signals from multiple scattering in the 
ambient, as well as the detector target material.  This would provide a unique opportunity to make direct contact with Planckian physics.  However, one still needs to provide an explanation of what 
kind of physics allows such heavy states to have significant interactions with the SM.  We pointed out that, generally speaking, such a signal could imply the presence of low mass, GeV-scale, messengers 
of Planck scale physics.  

We proposed that a ``dark'' $\ud$ gauge interaction could be a motivated GeV-scale mediator between SM and very large mass scales of $\ord{M_{\rm P}}$.  This is based on arguments that suggest $U(1)$ symmetries, which commonly arise in high scale theories, very often can survive spontaneous breaking, down to low scales, and can hence be good candidates for mediation between the SM and 
ultra heavy dark matter.  As a concrete example, we focused on $\ud$ coupled to baryon number, where the interactions of the associated vector boson $A_d$ of mass $\sim$ GeV with nucleons can provide the requisite values of the cross section for the detection of Planck scale dark matter, for phenomenologically allowed values of the $\ud$ coupling constant $\ad$.

The baryon current assumed in our work is anomalous and needs to be ultraviolet completed with 
additional fermions that carry electroweak charges, to cancel the anomalies.  These fermions then need to have masses $\gsim 100$~GeV to have escaped detection.  The range of parameters considered here suggests that the cutoff scale of the low energy effective theory can be 
$\sim 20$~TeV or higher and can accommodate such an ultraviolet completion.  However, the chiral assignment of  the fermions under $\ud$ implies that their masses are obtained from a Higgs field whose $\ud$ charge is small, in order to allow for a GeV-scale $A_d$.  
Hence, typically, our assumed model 
predicts several fermions at or above the weak scale, needed to cancel the anomalies.  If the masses of these fermions preserve electroweak symmetry, one generally expects longitudinally enhanced 
$A_d$ emissions in low energy processes.  This enhancement can lead to significant constraints on the parameter space of the model and can provide interesting low energy information on the underlying dynamics of the theory at or above the weak scale.

It may prove difficult to mediate direct detection of Planck scale DM, requiring substantial DM scattering cross sections against target material, assuming other vector interactions, such as gauged $B-L$, due to various stringent constraints on the gauge coupling constant, see for example Refs. \cite{Ilten:2018crw, Bauer:2018onh}.  In that case, the baryon current coupling discussed here could provide a well-motivated manner of linking sub-GeV and Planck scale physics.
We also note that other types of messengers, for example light scalars, can also potentially act as the GeV-Planck 
mediators, leading to different, and possibly less constrained, phenomenological and model building considerations.  In any event, due to the assumed non-negligible coupling between the low energy mediators and Planckian physics, questions about the stability 
of the assumed low masses against large quantum corrections could arise.  These questions may require assumptions about the non-trivial nature of $\ud$ symmetry breaking or composite 
mediators.  Hence, we posit that direct detection of Planckian DM would have interesting implications for ``naturalness'' versus  ``fine-tuning'' in physical theories.  

The assumed interactions can lead to kinetic mixing between the photon and $A_d$, parametrized by $\eps$.  However, in the mass range of interest to our analysis, we noted that ``dark photon'' constraints on $\eps$ do not apply directly, 
due to the dominance of the hadronic decays over leptonic decays of $A_d$.  For $\ad \sim 10^{-3}$, values of $\eps \lsim 10^{-2}$ may be allowed, as deduced by scaling of the current ``dark photon'' experimental constraints on leptonic decay modes.  
Future experiments, such as Belle II, can probe $A_d$ kinetic mixing in the leptonic channels, which 
could have interesting implications for the underlying dynamics of anomaly cancellation at or above the weak scale.  However, direct tests of the baryon current coupling to quarks would require searches that focus on the hadronic decays of $A_d$, for example into a pion and photon or to 3 pions. 
In addition to Belle II and other hadron factories, experiments such as GlueX may be highly sensitive to these types of decays \cite{Fanelli:2016utb}.
We also considered astrophysical constraints, related to the stability of white dwarfs against 
runaway nuclear fusion, on our model.  Those constraints, subject to some parametric assumptions,  typically disfavor DM masses above $\sim 10^{17}$~GeV in the context of our model, assuming potentially detectable DM-nucleon scattering.

\bigskip

\textbf{Acknowledgements.} We thank  Bogdan Dobrescu, Gordan Krnjaic, and Yue Zhang for very helpful discussions, and Jeff Dror for clarifications regarding the results in Refs.~\cite{Dror:2017ehi,Dror:2017nsg}. We would also like to thank Mike Williams for alerting us to further constraints and searches relevant to this study. This work is supported by the United States Department of Energy under Grant Contract DE-SC0012704. 

\bibliography{PSDMrefs}

\begin{thebibliography}{48}
\expandafter\ifx\csname natexlab\endcsname\relax\def\natexlab#1{#1}\fi
\expandafter\ifx\csname bibnamefont\endcsname\relax
  \def\bibnamefont#1{#1}\fi
\expandafter\ifx\csname bibfnamefont\endcsname\relax
  \def\bibfnamefont#1{#1}\fi
\expandafter\ifx\csname citenamefont\endcsname\relax
  \def\citenamefont#1{#1}\fi
\expandafter\ifx\csname url\endcsname\relax
  \def\url#1{\texttt{#1}}\fi
\expandafter\ifx\csname urlprefix\endcsname\relax\def\urlprefix{URL }\fi
\providecommand{\bibinfo}[2]{#2}
\providecommand{\eprint}[2][]{\url{#2}}

\bibitem[{\citenamefont{Bramante et~al.}(2018)\citenamefont{Bramante, Broerman,
  Lang, and Raj}}]{Bramante:2018qbc}
\bibinfo{author}{\bibfnamefont{J.}~\bibnamefont{Bramante}},
  \bibinfo{author}{\bibfnamefont{B.}~\bibnamefont{Broerman}},
  \bibinfo{author}{\bibfnamefont{R.~F.} \bibnamefont{Lang}}, \bibnamefont{and}
  \bibinfo{author}{\bibfnamefont{N.}~\bibnamefont{Raj}},
  \bibinfo{journal}{Phys. Rev.} \textbf{\bibinfo{volume}{D98}},
  \bibinfo{pages}{083516} (\bibinfo{year}{2018}), \eprint{1803.08044}.

\bibitem[{\citenamefont{Graham et~al.}(2018)\citenamefont{Graham, Janish,
  Narayan, Rajendran, and Riggins}}]{Graham:2018efk}
\bibinfo{author}{\bibfnamefont{P.~W.} \bibnamefont{Graham}},
  \bibinfo{author}{\bibfnamefont{R.}~\bibnamefont{Janish}},
  \bibinfo{author}{\bibfnamefont{V.}~\bibnamefont{Narayan}},
  \bibinfo{author}{\bibfnamefont{S.}~\bibnamefont{Rajendran}},
  \bibnamefont{and} \bibinfo{author}{\bibfnamefont{P.}~\bibnamefont{Riggins}}
  (\bibinfo{year}{2018}), \eprint{1805.07381}.

\bibitem[{\citenamefont{Halverson and Langacker}(2018)}]{Halverson:2018xge}
\bibinfo{author}{\bibfnamefont{J.}~\bibnamefont{Halverson}} \bibnamefont{and}
  \bibinfo{author}{\bibfnamefont{P.}~\bibnamefont{Langacker}},
  \bibinfo{journal}{PoS} \textbf{\bibinfo{volume}{TASI2017}},
  \bibinfo{pages}{019} (\bibinfo{year}{2018}), \eprint{1801.03503}.

\bibitem[{\citenamefont{Rajpoot}(1989)}]{Rajpoot:1989jb}
\bibinfo{author}{\bibfnamefont{S.}~\bibnamefont{Rajpoot}},
  \bibinfo{journal}{Phys. Rev.} \textbf{\bibinfo{volume}{D40}},
  \bibinfo{pages}{2421} (\bibinfo{year}{1989}).

\bibitem[{\citenamefont{Nelson and Tetradis}(1989)}]{Nelson:1989fx}
\bibinfo{author}{\bibfnamefont{A.~E.} \bibnamefont{Nelson}} \bibnamefont{and}
  \bibinfo{author}{\bibfnamefont{N.}~\bibnamefont{Tetradis}},
  \bibinfo{journal}{Phys. Lett.} \textbf{\bibinfo{volume}{B221}},
  \bibinfo{pages}{80} (\bibinfo{year}{1989}).

\bibitem[{\citenamefont{He and Rajpoot}(1990)}]{He:1989mi}
\bibinfo{author}{\bibfnamefont{X.-G.} \bibnamefont{He}} \bibnamefont{and}
  \bibinfo{author}{\bibfnamefont{S.}~\bibnamefont{Rajpoot}},
  \bibinfo{journal}{Phys. Rev.} \textbf{\bibinfo{volume}{D41}},
  \bibinfo{pages}{1636} (\bibinfo{year}{1990}).

\bibitem[{\citenamefont{Foot et~al.}(1989)\citenamefont{Foot, Joshi, and
  Lew}}]{Foot:1989ts}
\bibinfo{author}{\bibfnamefont{R.}~\bibnamefont{Foot}},
  \bibinfo{author}{\bibfnamefont{G.~C.} \bibnamefont{Joshi}}, \bibnamefont{and}
  \bibinfo{author}{\bibfnamefont{H.}~\bibnamefont{Lew}},
  \bibinfo{journal}{Phys. Rev.} \textbf{\bibinfo{volume}{D40}},
  \bibinfo{pages}{2487} (\bibinfo{year}{1989}).

\bibitem[{\citenamefont{Bailey and Davidson}(1995)}]{Bailey:1994qv}
\bibinfo{author}{\bibfnamefont{D.~C.} \bibnamefont{Bailey}} \bibnamefont{and}
  \bibinfo{author}{\bibfnamefont{S.}~\bibnamefont{Davidson}},
  \bibinfo{journal}{Phys. Lett.} \textbf{\bibinfo{volume}{B348}},
  \bibinfo{pages}{185} (\bibinfo{year}{1995}), \eprint{hep-ph/9411355}.

\bibitem[{\citenamefont{Carone and Murayama}(1995)}]{Carone:1994aa}
\bibinfo{author}{\bibfnamefont{C.~D.} \bibnamefont{Carone}} \bibnamefont{and}
  \bibinfo{author}{\bibfnamefont{H.}~\bibnamefont{Murayama}},
  \bibinfo{journal}{Phys. Rev. Lett.} \textbf{\bibinfo{volume}{74}},
  \bibinfo{pages}{3122} (\bibinfo{year}{1995}), \eprint{hep-ph/9411256}.

\bibitem[{\citenamefont{Aranda and Carone}(1998)}]{Aranda:1998fr}
\bibinfo{author}{\bibfnamefont{A.}~\bibnamefont{Aranda}} \bibnamefont{and}
  \bibinfo{author}{\bibfnamefont{C.~D.} \bibnamefont{Carone}},
  \bibinfo{journal}{Phys. Lett.} \textbf{\bibinfo{volume}{B443}},
  \bibinfo{pages}{352} (\bibinfo{year}{1998}), \eprint{hep-ph/9809522}.

\bibitem[{\citenamefont{Fileviez~Perez and Wise}(2010)}]{FileviezPerez:2010gw}
\bibinfo{author}{\bibfnamefont{P.}~\bibnamefont{Fileviez~Perez}}
  \bibnamefont{and} \bibinfo{author}{\bibfnamefont{M.~B.} \bibnamefont{Wise}},
  \bibinfo{journal}{Phys. Rev.} \textbf{\bibinfo{volume}{D82}},
  \bibinfo{pages}{011901} (\bibinfo{year}{2010}), \bibinfo{note}{[Erratum:
  Phys. Rev.D82,079901(2010)]}, \eprint{1002.1754}.

\bibitem[{\citenamefont{Fileviez~Perez and Wise}(2011)}]{FileviezPerez:2011pt}
\bibinfo{author}{\bibfnamefont{P.}~\bibnamefont{Fileviez~Perez}}
  \bibnamefont{and} \bibinfo{author}{\bibfnamefont{M.~B.} \bibnamefont{Wise}},
  \bibinfo{journal}{JHEP} \textbf{\bibinfo{volume}{08}}, \bibinfo{pages}{068}
  (\bibinfo{year}{2011}), \eprint{1106.0343}.

\bibitem[{\citenamefont{Graesser et~al.}(2011)\citenamefont{Graesser,
  Shoemaker, and Vecchi}}]{Graesser:2011vj}
\bibinfo{author}{\bibfnamefont{M.~L.} \bibnamefont{Graesser}},
  \bibinfo{author}{\bibfnamefont{I.~M.} \bibnamefont{Shoemaker}},
  \bibnamefont{and} \bibinfo{author}{\bibfnamefont{L.}~\bibnamefont{Vecchi}}
  (\bibinfo{year}{2011}), \eprint{1107.2666}.

\bibitem[{\citenamefont{Pospelov}(2011)}]{Pospelov:2011ha}
\bibinfo{author}{\bibfnamefont{M.}~\bibnamefont{Pospelov}},
  \bibinfo{journal}{Phys. Rev.} \textbf{\bibinfo{volume}{D84}},
  \bibinfo{pages}{085008} (\bibinfo{year}{2011}), \eprint{1103.3261}.

\bibitem[{\citenamefont{Pospelov and Pradler}(2012)}]{Pospelov:2012gm}
\bibinfo{author}{\bibfnamefont{M.}~\bibnamefont{Pospelov}} \bibnamefont{and}
  \bibinfo{author}{\bibfnamefont{J.}~\bibnamefont{Pradler}},
  \bibinfo{journal}{Phys. Rev.} \textbf{\bibinfo{volume}{D85}},
  \bibinfo{pages}{113016} (\bibinfo{year}{2012}), \bibinfo{note}{[Erratum:
  Phys. Rev.D88,no.3,039904(2013)]}, \eprint{1203.0545}.

\bibitem[{\citenamefont{Dobrescu and Frugiuele}(2014)}]{Dobrescu:2014fca}
\bibinfo{author}{\bibfnamefont{B.~A.} \bibnamefont{Dobrescu}} \bibnamefont{and}
  \bibinfo{author}{\bibfnamefont{C.}~\bibnamefont{Frugiuele}},
  \bibinfo{journal}{Phys. Rev. Lett.} \textbf{\bibinfo{volume}{113}},
  \bibinfo{pages}{061801} (\bibinfo{year}{2014}), \eprint{1404.3947}.

\bibitem[{\citenamefont{Tulin}(2014)}]{Tulin:2014tya}
\bibinfo{author}{\bibfnamefont{S.}~\bibnamefont{Tulin}},
  \bibinfo{journal}{Phys. Rev.} \textbf{\bibinfo{volume}{D89}},
  \bibinfo{pages}{114008} (\bibinfo{year}{2014}), \eprint{1404.4370}.

\bibitem[{\citenamefont{Batell et~al.}(2014)\citenamefont{Batell, deNiverville,
  McKeen, Pospelov, and Ritz}}]{Batell:2014yra}
\bibinfo{author}{\bibfnamefont{B.}~\bibnamefont{Batell}},
  \bibinfo{author}{\bibfnamefont{P.}~\bibnamefont{deNiverville}},
  \bibinfo{author}{\bibfnamefont{D.}~\bibnamefont{McKeen}},
  \bibinfo{author}{\bibfnamefont{M.}~\bibnamefont{Pospelov}}, \bibnamefont{and}
  \bibinfo{author}{\bibfnamefont{A.}~\bibnamefont{Ritz}},
  \bibinfo{journal}{Phys. Rev.} \textbf{\bibinfo{volume}{D90}},
  \bibinfo{pages}{115014} (\bibinfo{year}{2014}), \eprint{1405.7049}.

\bibitem[{\citenamefont{Fileviez~Perez
  et~al.}(2014)\citenamefont{Fileviez~Perez, Ohmer, and Patel}}]{Perez:2014qfa}
\bibinfo{author}{\bibfnamefont{P.}~\bibnamefont{Fileviez~Perez}},
  \bibinfo{author}{\bibfnamefont{S.}~\bibnamefont{Ohmer}}, \bibnamefont{and}
  \bibinfo{author}{\bibfnamefont{H.~H.} \bibnamefont{Patel}},
  \bibinfo{journal}{Phys. Lett.} \textbf{\bibinfo{volume}{B735}},
  \bibinfo{pages}{283} (\bibinfo{year}{2014}), \eprint{1403.8029}.

\bibitem[{\citenamefont{Coloma et~al.}(2016)\citenamefont{Coloma, Dobrescu,
  Frugiuele, and Harnik}}]{Coloma:2015pih}
\bibinfo{author}{\bibfnamefont{P.}~\bibnamefont{Coloma}},
  \bibinfo{author}{\bibfnamefont{B.~A.} \bibnamefont{Dobrescu}},
  \bibinfo{author}{\bibfnamefont{C.}~\bibnamefont{Frugiuele}},
  \bibnamefont{and} \bibinfo{author}{\bibfnamefont{R.}~\bibnamefont{Harnik}},
  \bibinfo{journal}{JHEP} \textbf{\bibinfo{volume}{04}}, \bibinfo{pages}{047}
  (\bibinfo{year}{2016}), \eprint{1512.03852}.

\bibitem[{\citenamefont{Fileviez~Perez}(2015)}]{Perez:2015rza}
\bibinfo{author}{\bibfnamefont{P.}~\bibnamefont{Fileviez~Perez}},
  \bibinfo{journal}{Phys. Rept.} \textbf{\bibinfo{volume}{597}},
  \bibinfo{pages}{1} (\bibinfo{year}{2015}), \eprint{1501.01886}.

\bibitem[{\citenamefont{Bardeen}(1969)}]{Bardeen:1969md}
\bibinfo{author}{\bibfnamefont{W.~A.} \bibnamefont{Bardeen}},
  \bibinfo{journal}{Phys. Rev.} \textbf{\bibinfo{volume}{184}},
  \bibinfo{pages}{1848} (\bibinfo{year}{1969}).

\bibitem[{\citenamefont{Halverson et~al.}(2014)\citenamefont{Halverson,
  Orlofsky, and Pierce}}]{Halverson:2014nwa}
\bibinfo{author}{\bibfnamefont{J.}~\bibnamefont{Halverson}},
  \bibinfo{author}{\bibfnamefont{N.}~\bibnamefont{Orlofsky}}, \bibnamefont{and}
  \bibinfo{author}{\bibfnamefont{A.}~\bibnamefont{Pierce}},
  \bibinfo{journal}{Phys. Rev.} \textbf{\bibinfo{volume}{D90}},
  \bibinfo{pages}{015002} (\bibinfo{year}{2014}), \eprint{1403.1592}.

\bibitem[{\citenamefont{Heister et~al.}(2002)}]{Heister:2002mn}
\bibinfo{author}{\bibfnamefont{A.}~\bibnamefont{Heister}} \bibnamefont{et~al.}
  (\bibinfo{collaboration}{ALEPH}), \bibinfo{journal}{Phys. Lett.}
  \textbf{\bibinfo{volume}{B533}}, \bibinfo{pages}{223} (\bibinfo{year}{2002}),
  \eprint{hep-ex/0203020}.

\bibitem[{\citenamefont{Preskill}(1991)}]{Preskill:1990fr}
\bibinfo{author}{\bibfnamefont{J.}~\bibnamefont{Preskill}},
  \bibinfo{journal}{Annals Phys.} \textbf{\bibinfo{volume}{210}},
  \bibinfo{pages}{323} (\bibinfo{year}{1991}).

\bibitem[{\citenamefont{Dror et~al.}(2017{\natexlab{a}})\citenamefont{Dror,
  Lasenby, and Pospelov}}]{Dror:2017ehi}
\bibinfo{author}{\bibfnamefont{J.~A.} \bibnamefont{Dror}},
  \bibinfo{author}{\bibfnamefont{R.}~\bibnamefont{Lasenby}}, \bibnamefont{and}
  \bibinfo{author}{\bibfnamefont{M.}~\bibnamefont{Pospelov}},
  \bibinfo{journal}{Phys. Rev. Lett.} \textbf{\bibinfo{volume}{119}},
  \bibinfo{pages}{141803} (\bibinfo{year}{2017}{\natexlab{a}}),
  \eprint{1705.06726}.

\bibitem[{\citenamefont{Dror et~al.}(2017{\natexlab{b}})\citenamefont{Dror,
  Lasenby, and Pospelov}}]{Dror:2017nsg}
\bibinfo{author}{\bibfnamefont{J.~A.} \bibnamefont{Dror}},
  \bibinfo{author}{\bibfnamefont{R.}~\bibnamefont{Lasenby}}, \bibnamefont{and}
  \bibinfo{author}{\bibfnamefont{M.}~\bibnamefont{Pospelov}},
  \bibinfo{journal}{Phys. Rev.} \textbf{\bibinfo{volume}{D96}},
  \bibinfo{pages}{075036} (\bibinfo{year}{2017}{\natexlab{b}}),
  \eprint{1707.01503}.

\bibitem[{\citenamefont{D'Hoker and
  Farhi}(1984{\natexlab{a}})}]{DHoker:1984izu}
\bibinfo{author}{\bibfnamefont{E.}~\bibnamefont{D'Hoker}} \bibnamefont{and}
  \bibinfo{author}{\bibfnamefont{E.}~\bibnamefont{Farhi}},
  \bibinfo{journal}{Nucl. Phys.} \textbf{\bibinfo{volume}{B248}},
  \bibinfo{pages}{59} (\bibinfo{year}{1984}{\natexlab{a}}).

\bibitem[{\citenamefont{D'Hoker and
  Farhi}(1984{\natexlab{b}})}]{DHoker:1984mif}
\bibinfo{author}{\bibfnamefont{E.}~\bibnamefont{D'Hoker}} \bibnamefont{and}
  \bibinfo{author}{\bibfnamefont{E.}~\bibnamefont{Farhi}},
  \bibinfo{journal}{Nucl. Phys.} \textbf{\bibinfo{volume}{B248}},
  \bibinfo{pages}{77} (\bibinfo{year}{1984}{\natexlab{b}}).

\bibitem[{\citenamefont{Holdom}(1986)}]{Holdom:1985ag}
\bibinfo{author}{\bibfnamefont{B.}~\bibnamefont{Holdom}},
  \bibinfo{journal}{Phys. Lett.} \textbf{\bibinfo{volume}{166B}},
  \bibinfo{pages}{196} (\bibinfo{year}{1986}).

\bibitem[{\citenamefont{Gopalakrishna et~al.}(2008)\citenamefont{Gopalakrishna,
  Jung, and Wells}}]{Gopalakrishna:2008dv}
\bibinfo{author}{\bibfnamefont{S.}~\bibnamefont{Gopalakrishna}},
  \bibinfo{author}{\bibfnamefont{S.}~\bibnamefont{Jung}}, \bibnamefont{and}
  \bibinfo{author}{\bibfnamefont{J.~D.} \bibnamefont{Wells}},
  \bibinfo{journal}{Phys. Rev.} \textbf{\bibinfo{volume}{D78}},
  \bibinfo{pages}{055002} (\bibinfo{year}{2008}), \eprint{0801.3456}.

\bibitem[{\citenamefont{Pospelov et~al.}(2008)\citenamefont{Pospelov, Ritz, and
  Voloshin}}]{Pospelov:2007mp}
\bibinfo{author}{\bibfnamefont{M.}~\bibnamefont{Pospelov}},
  \bibinfo{author}{\bibfnamefont{A.}~\bibnamefont{Ritz}}, \bibnamefont{and}
  \bibinfo{author}{\bibfnamefont{M.~B.} \bibnamefont{Voloshin}},
  \bibinfo{journal}{Phys. Lett.} \textbf{\bibinfo{volume}{B662}},
  \bibinfo{pages}{53} (\bibinfo{year}{2008}), \eprint{0711.4866}.

\bibitem[{\citenamefont{Essig et~al.}(2012)\citenamefont{Essig, Mardon, and
  Volansky}}]{Essig:2011nj}
\bibinfo{author}{\bibfnamefont{R.}~\bibnamefont{Essig}},
  \bibinfo{author}{\bibfnamefont{J.}~\bibnamefont{Mardon}}, \bibnamefont{and}
  \bibinfo{author}{\bibfnamefont{T.}~\bibnamefont{Volansky}},
  \bibinfo{journal}{Phys. Rev.} \textbf{\bibinfo{volume}{D85}},
  \bibinfo{pages}{076007} (\bibinfo{year}{2012}), \eprint{1108.5383}.

\bibitem[{\citenamefont{Aaij et~al.}(2015)}]{Aaij:2015tna}
\bibinfo{author}{\bibfnamefont{R.}~\bibnamefont{Aaij}} \bibnamefont{et~al.}
  (\bibinfo{collaboration}{LHCb}), \bibinfo{journal}{Phys. Rev. Lett.}
  \textbf{\bibinfo{volume}{115}}, \bibinfo{pages}{161802}
  (\bibinfo{year}{2015}), \eprint{1508.04094}.

\bibitem[{\citenamefont{Aaij et~al.}(2017)}]{Aaij:2016qsm}
\bibinfo{author}{\bibfnamefont{R.}~\bibnamefont{Aaij}} \bibnamefont{et~al.}
  (\bibinfo{collaboration}{LHCb}), \bibinfo{journal}{Phys. Rev.}
  \textbf{\bibinfo{volume}{D95}}, \bibinfo{pages}{071101}
  (\bibinfo{year}{2017}), \eprint{1612.07818}.

\bibitem[{\citenamefont{Ilten et~al.}(2018)\citenamefont{Ilten, Soreq,
  Williams, and Xue}}]{Ilten:2018crw}
\bibinfo{author}{\bibfnamefont{P.}~\bibnamefont{Ilten}},
  \bibinfo{author}{\bibfnamefont{Y.}~\bibnamefont{Soreq}},
  \bibinfo{author}{\bibfnamefont{M.}~\bibnamefont{Williams}}, \bibnamefont{and}
  \bibinfo{author}{\bibfnamefont{W.}~\bibnamefont{Xue}},
  \bibinfo{journal}{JHEP} \textbf{\bibinfo{volume}{06}}, \bibinfo{pages}{004}
  (\bibinfo{year}{2018}), \eprint{1801.04847}.

\bibitem[{\citenamefont{Lees et~al.}(2014)}]{Lees:2014xha}
\bibinfo{author}{\bibfnamefont{J.~P.} \bibnamefont{Lees}} \bibnamefont{et~al.}
  (\bibinfo{collaboration}{BaBar}), \bibinfo{journal}{Phys. Rev. Lett.}
  \textbf{\bibinfo{volume}{113}}, \bibinfo{pages}{201801}
  (\bibinfo{year}{2014}), \eprint{1406.2980}.

\bibitem[{\citenamefont{Babusci et~al.}(2014)}]{Babusci:2014sta}
\bibinfo{author}{\bibfnamefont{D.}~\bibnamefont{Babusci}} \bibnamefont{et~al.}
  (\bibinfo{collaboration}{KLOE-2}), \bibinfo{journal}{Phys. Lett.}
  \textbf{\bibinfo{volume}{B736}}, \bibinfo{pages}{459} (\bibinfo{year}{2014}),
  \eprint{1404.7772}.

\bibitem[{\citenamefont{Anastasi et~al.}(2016)}]{Anastasi:2016ktq}
\bibinfo{author}{\bibfnamefont{A.}~\bibnamefont{Anastasi}} \bibnamefont{et~al.}
  (\bibinfo{collaboration}{KLOE-2}), \bibinfo{journal}{Phys. Lett.}
  \textbf{\bibinfo{volume}{B757}}, \bibinfo{pages}{356} (\bibinfo{year}{2016}),
  \eprint{1603.06086}.

\bibitem[{\citenamefont{Abe et~al.}(2010)}]{Abe:2010gxa}
\bibinfo{author}{\bibfnamefont{T.}~\bibnamefont{Abe}} \bibnamefont{et~al.}
  (\bibinfo{collaboration}{Belle-II}) (\bibinfo{year}{2010}),
  \eprint{1011.0352}.

\bibitem[{\citenamefont{Kou et~al.}(2018)}]{Kou:2018nap}
\bibinfo{author}{\bibfnamefont{E.}~\bibnamefont{Kou}} \bibnamefont{et~al.}
  (\bibinfo{year}{2018}), \eprint{1808.10567}.

\bibitem[{\citenamefont{Patrignani et~al.}(2016)}]{Patrignani:2016xqp}
\bibinfo{author}{\bibfnamefont{C.}~\bibnamefont{Patrignani}}
  \bibnamefont{et~al.} (\bibinfo{collaboration}{Particle Data Group}),
  \bibinfo{journal}{Chin. Phys.} \textbf{\bibinfo{volume}{C40}},
  \bibinfo{pages}{100001} (\bibinfo{year}{2016}).

\bibitem[{\citenamefont{Hook et~al.}(2011)\citenamefont{Hook, Izaguirre, and
  Wacker}}]{Hook:2010tw}
\bibinfo{author}{\bibfnamefont{A.}~\bibnamefont{Hook}},
  \bibinfo{author}{\bibfnamefont{E.}~\bibnamefont{Izaguirre}},
  \bibnamefont{and} \bibinfo{author}{\bibfnamefont{J.~G.}
  \bibnamefont{Wacker}}, \bibinfo{journal}{Adv. High Energy Phys.}
  \textbf{\bibinfo{volume}{2011}}, \bibinfo{pages}{859762}
  (\bibinfo{year}{2011}), \eprint{1006.0973}.

\bibitem[{\citenamefont{Curtin et~al.}(2015)\citenamefont{Curtin, Essig, Gori,
  and Shelton}}]{Curtin:2014cca}
\bibinfo{author}{\bibfnamefont{D.}~\bibnamefont{Curtin}},
  \bibinfo{author}{\bibfnamefont{R.}~\bibnamefont{Essig}},
  \bibinfo{author}{\bibfnamefont{S.}~\bibnamefont{Gori}}, \bibnamefont{and}
  \bibinfo{author}{\bibfnamefont{J.}~\bibnamefont{Shelton}},
  \bibinfo{journal}{JHEP} \textbf{\bibinfo{volume}{02}}, \bibinfo{pages}{157}
  (\bibinfo{year}{2015}), \eprint{1412.0018}.

\bibitem[{\citenamefont{Bennett et~al.}(2006)}]{Bennett:2006fi}
\bibinfo{author}{\bibfnamefont{G.~W.} \bibnamefont{Bennett}}
  \bibnamefont{et~al.} (\bibinfo{collaboration}{Muon g-2}),
  \bibinfo{journal}{Phys. Rev.} \textbf{\bibinfo{volume}{D73}},
  \bibinfo{pages}{072003} (\bibinfo{year}{2006}), \eprint{hep-ex/0602035}.

\bibitem[{\citenamefont{Pospelov}(2009)}]{Pospelov:2008zw}
\bibinfo{author}{\bibfnamefont{M.}~\bibnamefont{Pospelov}},
  \bibinfo{journal}{Phys. Rev.} \textbf{\bibinfo{volume}{D80}},
  \bibinfo{pages}{095002} (\bibinfo{year}{2009}), \eprint{0811.1030}.

\bibitem[{\citenamefont{Bauer et~al.}(2018)\citenamefont{Bauer, Foldenauer, and
  Jaeckel}}]{Bauer:2018onh}
\bibinfo{author}{\bibfnamefont{M.}~\bibnamefont{Bauer}},
  \bibinfo{author}{\bibfnamefont{P.}~\bibnamefont{Foldenauer}},
  \bibnamefont{and} \bibinfo{author}{\bibfnamefont{J.}~\bibnamefont{Jaeckel}},
  \bibinfo{journal}{JHEP} \textbf{\bibinfo{volume}{07}}, \bibinfo{pages}{094}
  (\bibinfo{year}{2018}), \eprint{1803.05466}.

\bibitem[{\citenamefont{Fanelli and Williams}(2017)}]{Fanelli:2016utb}
\bibinfo{author}{\bibfnamefont{C.}~\bibnamefont{Fanelli}} \bibnamefont{and}
  \bibinfo{author}{\bibfnamefont{M.}~\bibnamefont{Williams}},
  \bibinfo{journal}{J. Phys.} \textbf{\bibinfo{volume}{G44}},
  \bibinfo{pages}{014002} (\bibinfo{year}{2017}), \eprint{1605.07161}.

\end{thebibliography}
\end{document}